\RequirePackage{fix-cm}
\documentclass[smallextended]{svjour3}       
\smartqed  
\usepackage{graphicx}
\usepackage{color}
\begin{document}

\title{Multiple rogue wave solutions for (2+1)-dimensional Boussinesq equation\thanks{Project supported by Science and Technology Project of Education Department
of Jiangxi Province(GJJ151079).}}


\author{ Jian-Guo Liu$^{1*}$, Wen-Hui Zhu$^{2}$
}


\institute{Jian-Guo Liu(*Corresponding author)\at
              College of Computer, Jiangxi University of
Traditional Chinese Medicine, Jiangxi 330004, China\\
              Tel.: +8613970042436\\
              \email{20101059@jxutcm.edu.cn}\\
\and  Wen-Hui Zhu \at Institute of artificial intelligence, Nanchang
Institute of Science and Technology, Jiangxi 330108, China\\
              \email{415422402@qq.com}}

\date{Received: date / Accepted: date}

\maketitle

\begin{abstract}
In this paper, a modified symbolic computation approach is proposed.
The multiple rogue wave solutions of a generalized (2+1)-dimensional
Boussinesq equation are obtained by using the modified symbolic
computation approach. Dynamics features of these obtained multiple
rogue wave solutions are displayed in 3D and contour plots. Compared
with the original symbolic computation approach, our method does not
need to find Hirota bilinear form of nonlinear system.

\keywords{Rogue wave, modified symbolic computation approach,
(2+1)-dimensional Boussinesq equation.}
\subclass{35C08 \and 68M07 \and 33F10}
\end{abstract}

\section{Introduction}
\label{intro} \quad Many important natural sciences and engineering
problems can be attributed to the study of the nonlinear partial
differential equations (NPDEs) [1-3]. Therefore, the research on the
exact solutions of the NPDEs has shown a very important theory and
application value. At present, it has not been provided a universal
and effective method for obtaining the exact solution [4-8]. Since
the 1950s, people put forward the concept of ``soliton" in the study
of nonlinear phenomena, which  becomes a hot topic in nonlinear
science [9-13]. Recently, rogue wave (a special soliton) attracts a
lot of attention and appears in many important areas, such as
optical fibres, Bose-Einstein condensates, super fluids and so on
[14]. Rogue wave solutions of many NPDEs have been investigated
based the Hirota bilinear method [15-18]. These studies mainly focus
on  obtaining the low order rogue wave solutions. Multiple rogue
waves are difficult to get, so there are few literatures on it. It
is inspired by the symbolic computation approach proposed by Zha
[19], we present a modified symbolic computation approach for
studying the multiple rogue wave solutions of NPDEs.

\quad In this paper, we investigate the following generalized
(2+1)-dimensional Boussinesq equation [20]
\begin{eqnarray}
\alpha  \left(2 u_x^2+2 u
   u_{xx}\right)+\beta  u_{xxxx}+\gamma
   u_{xx}+u_{tt}-u_{yy}=0,
\end{eqnarray}where $u=u(x,y,t)$, $\alpha$, $\beta$ and $\gamma$ are arbitrary real
constants. Eq. (1) represents the propagation of gravity waves on
the earth's surface. Especially the oncoming collision of the
oblique wave. When $\alpha, \beta$ and $\gamma$ choose different
values, Eq (1) can be reduced to (1 + 1)-dimensional Boussinesq
equation [21], classical Benjamin-Ono equation [22], (2 + 1)-
dimensional Benjamin-Ono equation [23] and (2+1)-dimensional
Boussinesq water equation [24]. The breathers and first-order rogue
wave solutions have been obtained in Ref. [25] when $\alpha=3
\beta$. When $\alpha=\gamma=-1, \beta=-3$, the first-order lump
solutions are studied in Ref. [26]. Besides, the high-order breather
and lump solutions have been also presented in Ref. [20]. However,
multiple rogue wave solutions have not been studied by using the
symbolic computation approach.

\quad The organization of this paper is as follows. Section 2
proposes a modified symbolic computation approach; Section 3 obtains
the 1-rogue wave solutions; Section 4 presents the 3-rogue wave
solutions; Section 5 studies the 6-rogue wave solutions; Section 6
makes this conclusions.

\section{Modified symbolic computation approach} \label{sec:2}
\quad In Ref.[19], a symbolic computation approach was proposed by
Zha. Multiple rogue wave solutions of many NPDEs have been discussed
by using the symbolic computation approach [27-31]. Here,
we want to make a slight adjustment to this method as follows\\

Step1. Make a traveling wave transformation $\upsilon=x-\omega t$ in
the following (2+1)-dimensional nonlinear system
{\begin{eqnarray}\Xi(u, u_t, u_x, u_y, u_{xy},
\cdots)=0,\end{eqnarray} }Eq. (2) will become a (1+1)-dimensional
equation about $\upsilon$ and $y$ {\begin{eqnarray}\Xi(u,
u_\upsilon, u_y, u_{\upsilon y}, \cdots)=0.
\end{eqnarray} }
Step2. By Painlev\'{e} analysis, we make the following
transformation {\begin{eqnarray}u(\upsilon,
y)=u_0+\frac{\partial^n}{\partial\upsilon^m}ln\xi(\upsilon, y).
\end{eqnarray} }Substituting Eq. (4) into Eq. (3) and balancing the order of the highest derivative term and nonlinear
term, $m$ can be obtained. We point out that the multiple rogue wave
solutions of Eq. (3) can be obtained without deriving the Hirota
bilinear form, but the Hirota bilinear form is needed in the
original method.

Step3. Supposing {\begin{eqnarray}\xi(\upsilon, y)=F_{n+1}(\upsilon,
y)+2 \nu y P_n(\upsilon, y)+2 \mu \upsilon Q_n(\upsilon,
y)+(\mu^2+\nu^2) F_{n-1}(\upsilon, y),
\end{eqnarray} }with
{\begin{eqnarray}F_n(\upsilon, y)&=&\sum^{n(n+1)/2}_{k=0}\sum^{k}_{i=0}a_{n(n+1)-2k,2i}y^{2i}\upsilon^{n(n+1)-2k},\nonumber\\
 P_n(\upsilon, y)&=&\sum^{n(n+1)/2}_{k=0}\sum^{k}_{i=0}b_{n(n+1)-2k,2i}\upsilon^{2i}y^{n(n+1)-2k},\nonumber\\
  Q_n(\upsilon, y)&=&\sum^{n(n+1)/2}_{k=0}\sum^{k}_{i=0}c_{n(n+1)-2k,2i}y^{2i}\upsilon^{n(n+1)-2k},\nonumber
\end{eqnarray} }$F_0=1, F_{-1}=P_0=Q_0=0$, where $a_{m,l}$,
$b_{m,l}$ and $c_{m,l}$($m, l\in {[0, 2,4, \cdots, n(n+1)]}$) are
unknown constants, $\mu$ and $\nu$ are the wave center.

Step4. Substituting Eq. (4) and Eq. (5) into Eq. (3) and equating
all the coefficients of the different powers of $\upsilon$ and $y$
to zero, the values of $a_{m,l}$, $b_{m,l}$ and $c_{m,l}$($m, l\in
{[0, 2,4, \cdots, n(n+1)]}$) can be obtained. Substituting these
values into Eq. (4) and Eq. (5), the corresponding multiple rogue
wave solutions of Eq. (2) are derived.

\section{1-rogue wave solutions} \label{sec:2}
\quad Based on previous literature and modified symbolic computation
approach [27-31], assume {\begin{eqnarray}\upsilon=x-\omega t,
u=\frac{6
\beta}{\alpha}\,[ln\xi(\upsilon,y)]_{\upsilon\upsilon},\end{eqnarray}
}Eq. (1) can be changed as{\begin{eqnarray} &&\xi^3 [\beta
\xi_{\upsilon\upsilon\upsilon\upsilon\upsilon\upsilon}+\left(\gamma
   +\omega ^2\right) \xi_{\upsilon\upsilon\upsilon\upsilon}-\xi_{\upsilon\upsilon yy}]+\xi^2 [\xi_{\upsilon\upsilon} [-3
   \beta  \xi_{\upsilon\upsilon\upsilon\upsilon}-3 \left(\gamma +\omega ^2\right) \xi_{\upsilon\upsilon}\nonumber\\&&+\xi_{yy}]
   +2 \xi_\upsilon \left(-3 \beta  \xi_{\upsilon\upsilon\upsilon\upsilon\upsilon}-2
   \left(\gamma +\omega ^2\right) \xi_{\upsilon\upsilon\upsilon}+\xi_{\upsilon yy}\right)+2 \beta  \xi_{\upsilon\upsilon\upsilon}^2\nonumber\\&&
   +2 \xi_{\upsilon y}^2]+2 \xi  [\xi_\upsilon^2 [9 \beta  \xi_{\upsilon\upsilon\upsilon\upsilon}+6
   \left(\gamma +\omega ^2\right) \xi_{\upsilon\upsilon}]-3 \beta
   \xi_{\upsilon\upsilon}^3-\xi_{yy} \xi_\upsilon^2]\nonumber\\&&-6 \xi_\upsilon^2 [4
   \beta  \xi_{\upsilon\upsilon\upsilon} \xi_\upsilon-3 \beta
     \xi_{\upsilon\upsilon}^2+\left(\gamma +\omega ^2\right) \xi_\upsilon^2]+2 \xi_y \left(\xi \xi_{\upsilon\upsilon y}
     -4 \xi_\upsilon \xi_{\upsilon y}\right) \xi\nonumber\\&&+\xi_y^2 \left(6 \xi_\upsilon^2-2 \xi  \xi_{\upsilon\upsilon}\right).
\end{eqnarray}
}In here, we don't need to find the Hirota bilinear form of Eq. (7).
In Eq. (7), we select {\begin{eqnarray} \xi(\upsilon,y)=(\upsilon
-\mu )^2+\zeta _1 (y-\nu )^2+\zeta _0,
\end{eqnarray}
}where $\mu$, $\nu$, $\zeta _0$ and $\zeta _1$ are undetermined real
constants. Substituting Eq. (8) into Eq. (7) and equating the
coefficients of all powers $\upsilon$ and $y$ to zero, we obtain
\begin{eqnarray}
\zeta_0=-\frac{3 \beta }{\gamma +\omega ^2}, \zeta_1=-\gamma -\omega
^2.
\end{eqnarray}Substituting Eq. (8) and Eq. (8) into Eq. (6), the 1-rogue wave
solutions for Eq. (1) can be got as {\begin{eqnarray} u=\frac{12
\beta  [-\frac{3 \beta }{\gamma +\omega ^2}-(\mu -\upsilon
   )^2-\left(\gamma +\omega ^2\right) (y-\nu )^2]}{\alpha
   [-\frac{3 \beta }{\gamma +\omega ^2}+(\mu -\upsilon )^2-\left(\gamma
   +\omega ^2\right) (y-\nu )^2]^2}.
\end{eqnarray}}Rogue wave
(10) has three extreme value points $(\mu, \nu)$, $(\mu \pm\frac{3
\beta }{\sqrt{-\beta  \left(\gamma +\omega ^2\right)}}, \nu)$.  The
maximum point $(\mu, \nu)$ is the peak of the rogue wave, and  the
minimum points $(\mu \pm\frac{3 \beta }{\sqrt{-\beta  \left(\gamma
+\omega ^2\right)}}, \nu)$ are two valleys. The centers of the peak
and valleys  are located on a straight line $y =\nu$, and the two
valleys are symmetric with respect to the peak. Fig. 1 shows the
dynamics features of rogue wave (10).

\includegraphics[scale=0.55,bb=-20 270 10 10]{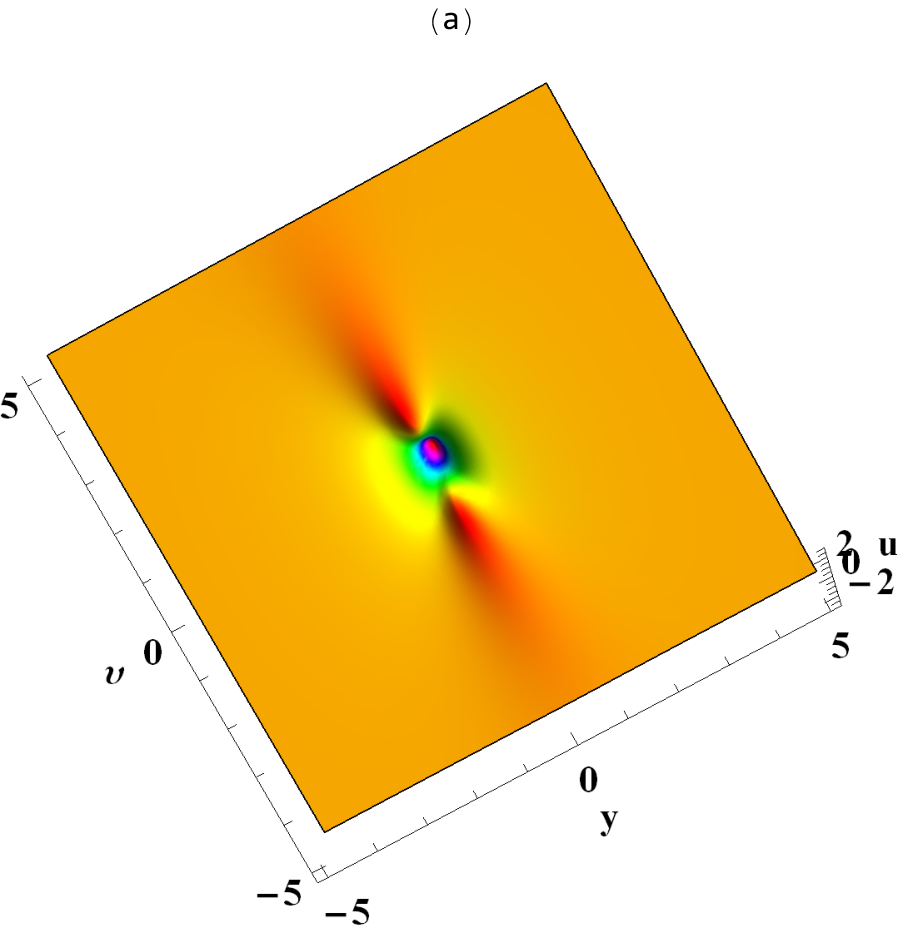}
\includegraphics[scale=0.45,bb=-270 310 10 10]{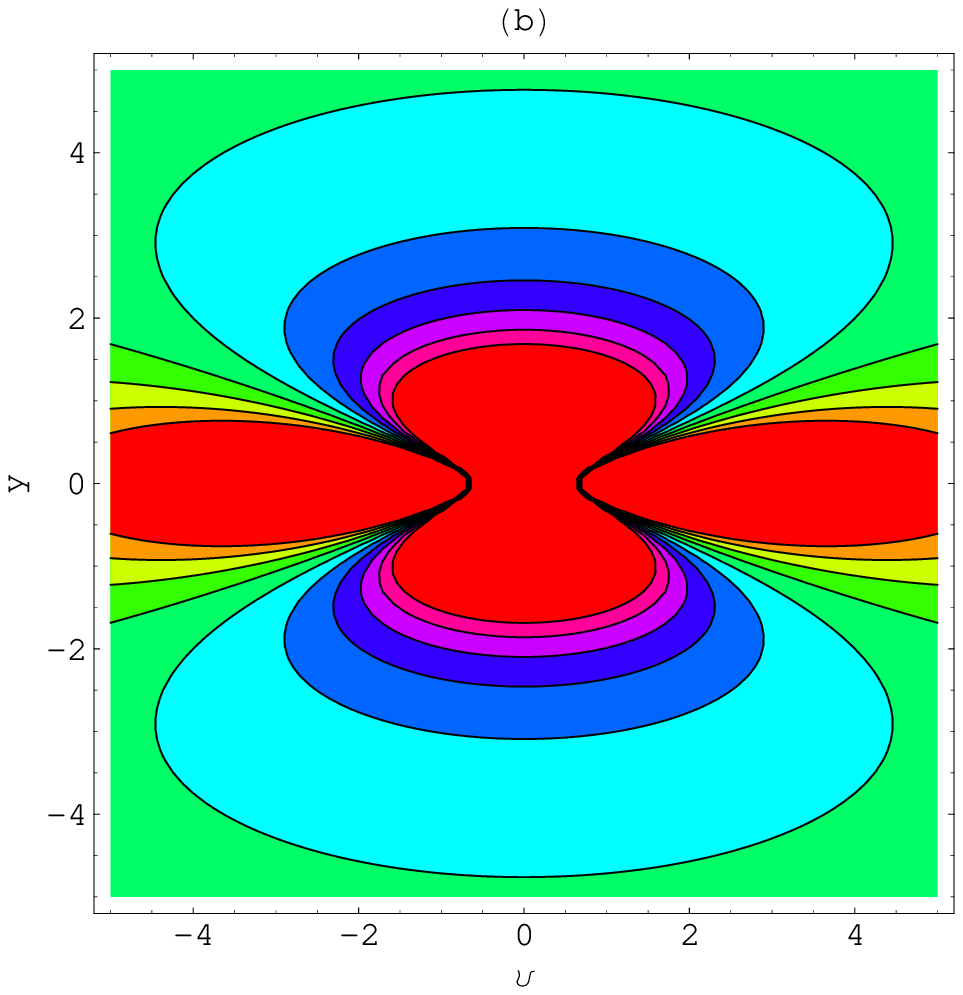}
\vspace{5.5cm}
\begin{tabbing}
\textbf{Fig. 1}. Rogue wave (10) with $\mu=\nu=0$, $\alpha=12$,
$\gamma=-8$, $\beta=\omega=1$,\\ (a) 3D graphic,  (b) contour plot.
\end{tabbing}

\includegraphics[scale=0.55,bb=-20 270 10 10]{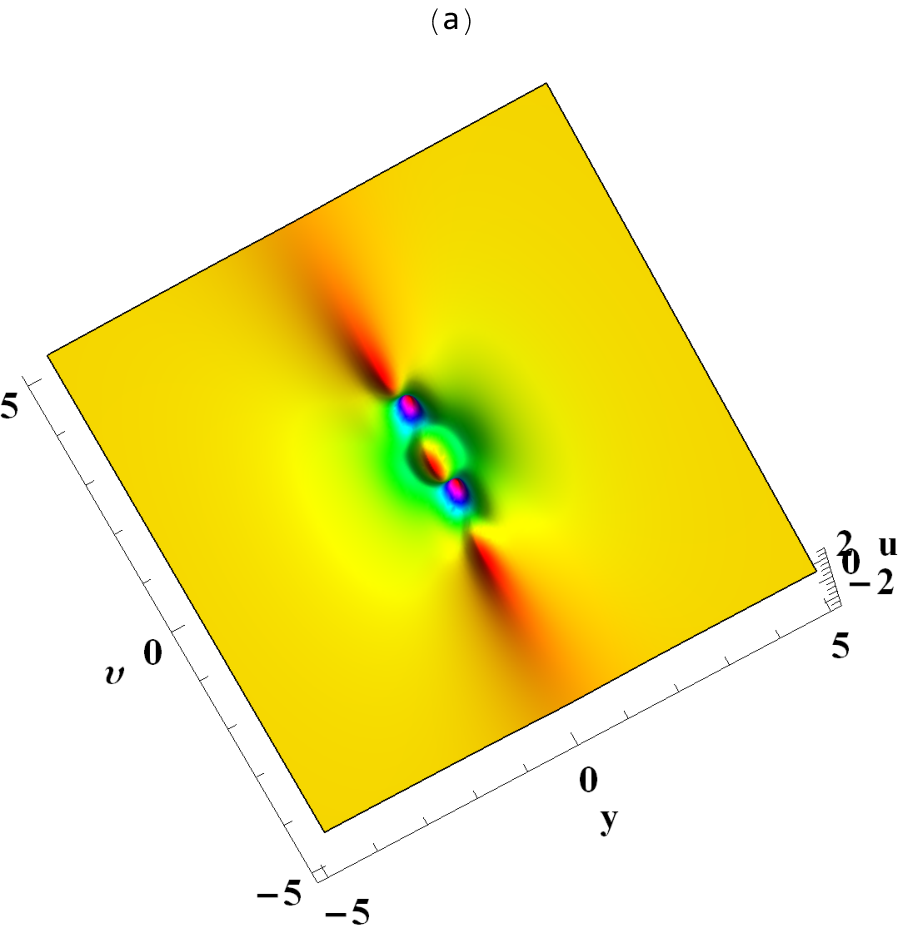}
\includegraphics[scale=0.45,bb=-270 310 10 10]{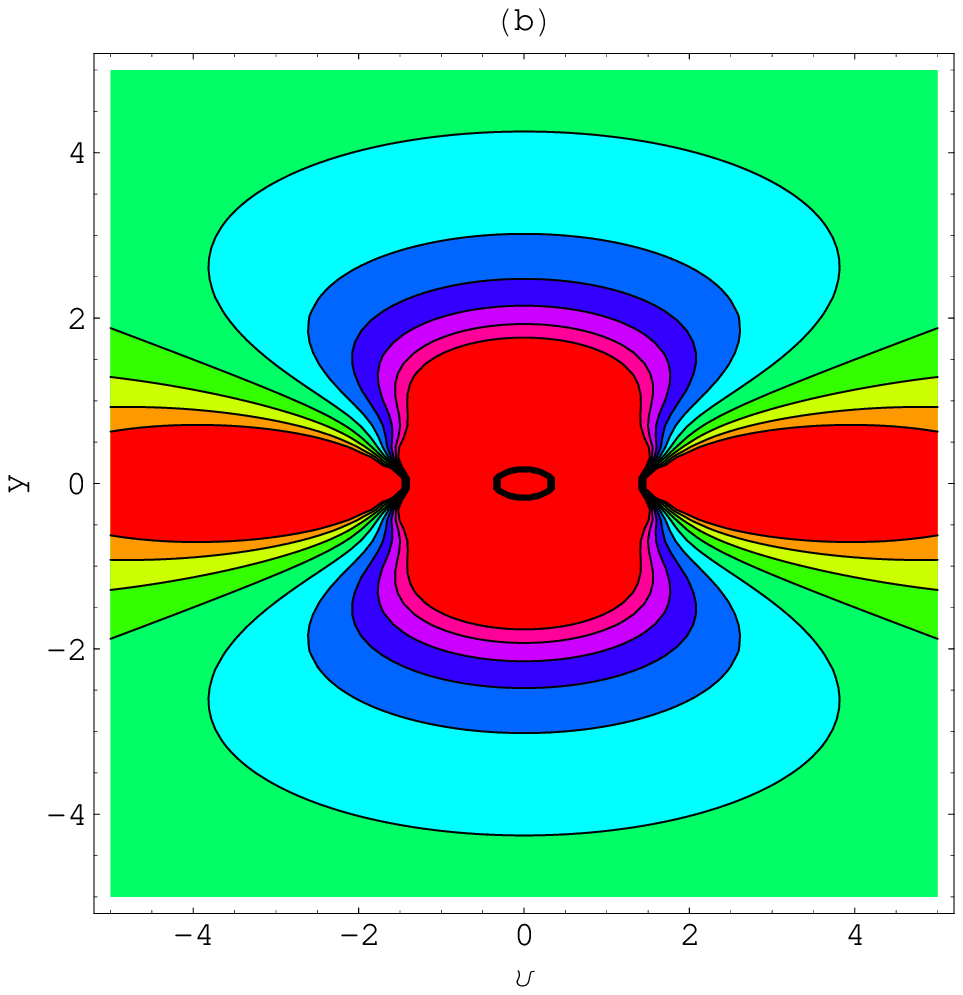}
\vspace{5.5cm}
\begin{tabbing}
\textbf{Fig. 2}. Rogue wave (13) with  $\alpha=12$, $\gamma=-8$,
$\beta=\omega=\zeta_{21}=\zeta_{24}=1$,\\ $\mu=\nu=0$, (a) 3D
graphic, (b) contour plot.
\end{tabbing}

\section{3-rogue wave solutions} \label{sec:2}
In order to look for obtaining the 3-rogue wave solutions, we choose
{\begin{eqnarray} \xi(\upsilon,y)&=&\mu ^2+\nu ^2+\upsilon ^6+y^6
\zeta _{17}+y^4 \zeta _{16}+2 \mu  \upsilon  \left(y^2
   \zeta _{23}+\upsilon ^2 \zeta _{24}+\zeta _{22}\right)\nonumber\\&+&2 \nu  y \left(y^2 \zeta
   _{20}+\upsilon ^2 \zeta _{21}+\zeta _{19}\right)+\upsilon ^4 y^2 \zeta _{11}+y^2
   \zeta _{15}\nonumber\\&+&\upsilon ^2 \left(y^4 \zeta _{14}+y^2 \zeta _{13}+\zeta
   _{12}\right)+\upsilon ^4 \zeta _{10}+\zeta _{18},
\end{eqnarray}
}where $\zeta _i (i=10,\cdots, 24)$  is undetermined real constant.
Substituting Eq. (11) into Eq. (7) and equating the coefficients of
all powers $\upsilon$ and $y$ to zero, we derive
\begin{eqnarray}
\zeta_{11}&=&-3 \left(\gamma +\omega ^2\right), \zeta_{14}=3
\left(\gamma +\omega ^2\right)^2, \zeta_{16}=-17 \beta  \left(\gamma +\omega ^2\right),\nonumber\\
\zeta_{13}&=&90 \beta, \zeta_{20}=\frac{1}{3} \zeta
_{21} \left(\gamma +\omega ^2\right), \zeta_{17}=-\left(\gamma +\omega ^2\right)^3,\nonumber\\
\zeta_{15}&=&-\frac{475 \beta ^2}{\gamma +\omega ^2}, \zeta_{23}=3
\zeta _{24} \left(\gamma +\omega ^2\right),
\zeta_{22}=\frac{\beta  \zeta _{24}}{\gamma +\omega ^2},\nonumber\\
\zeta_{12}&=&-\frac{125 \beta ^2}{\left(\gamma +\omega ^2\right)^2},
\zeta_{10}=-\frac{25 \beta }{\gamma +\omega ^2}, \zeta_{19}=-\frac{5 \beta  \zeta _{21}}{3 \left(\gamma +\omega ^2\right)},\nonumber\\
\zeta_{18}&=&-[9 [1875 \beta ^3+\left(\gamma +\omega ^2\right)^3
\left(\mu
   ^2+\nu ^2\right)-\mu ^2 \zeta _{24}^2 \left(\gamma +\omega
   ^2\right)^3]\nonumber\\&+&\nu ^2 \zeta _{21}^2 \left(\gamma +\omega
   ^2\right)^2]/[9 \left(\gamma +\omega ^2\right)^3].
\end{eqnarray}Substituting Eq. (11) and Eq. (12) into Eq. (6), the 3-rogue wave
solutions for Eq. (1) can be written as {\begin{eqnarray} u&=&[486
\beta \left(\gamma +\omega ^2\right)^6 [[2 [5
   \left(-\frac{25 \beta ^2}{\left(\gamma +\omega ^2\right)^2}-\frac{30
   \beta  \upsilon ^2}{\gamma +\omega ^2}+3 \upsilon ^4\right)+3 y^4
   \left(\gamma +\omega ^2\right)^2\nonumber\\&+&18 y^2 [5 \beta -\upsilon ^2
   \left(\gamma +\omega ^2\right)]+2 \nu  y \zeta _{21}+6 \mu
   \upsilon  \zeta _{24}] [9 [-1875 \beta ^3\nonumber\\&-&25 \beta ^2
   \left(\gamma +\omega ^2\right) [5 \upsilon ^2+19 y^2 \left(\gamma
   +\omega ^2\right)]-\beta  \left(\gamma +\omega ^2\right)^2
   [y^2 \left(\gamma +\omega ^2\right)-5 \upsilon ^2]\nonumber\\&*&[17
   y^2 \left(\gamma +\omega ^2\right)-5 \upsilon ^2]-\left(\gamma
   +\omega ^2\right)^3 [y^2 \left(\gamma +\omega ^2\right)-\upsilon
   ^2]^3]\nonumber\\&+&\left(\gamma +\omega ^2\right)^2 [9 \mu
   \zeta _{24} [\mu  \zeta _{24} \left(\gamma +\omega
   ^2\right)+2 \upsilon  [\beta +\left(\gamma +\omega ^2\right)
   \left(\upsilon ^2+3 y^2 \left(\gamma +\omega
   ^2\right)\right)]]\nonumber\\&+&6 \nu  y \zeta _{21}
   [\left(\gamma +\omega ^2\right) [3 \upsilon ^2+y^2 \left(\gamma
   +\omega ^2\right)]-5 \beta ]-\nu ^2 \zeta
   _{21}^2]]]/[9 \left(\gamma +\omega ^2\right)^3]\nonumber\\&-&4
   [[\upsilon  [-125 \beta ^2+10 \beta  \left(\gamma +\omega
   ^2\right) \left(9 y^2 \left(\gamma +\omega ^2\right)-5 \upsilon
   ^2\right)+3 \left(\gamma +\omega ^2\right)^2 [\upsilon ^2\nonumber\\&-&y^2
   \left(\gamma +\omega ^2\right)]^2]+\mu  \zeta _{24}
   \left(\gamma +\omega ^2\right) [\beta +3 \left(\gamma +\omega
   ^2\right) \left(\upsilon ^2+y^2 \left(\gamma +\omega
   ^2\right)\right)]]\nonumber\\&/&[\left(\gamma +\omega ^2\right)^2]+2 \nu
   \upsilon  y \zeta _{21}]{}^2]]/[\alpha  [9
   [-1875 \beta ^3-25 \beta ^2 \left(\gamma +\omega ^2\right) [5
   \upsilon ^2\nonumber\\&+&19 y^2 \left(\gamma +\omega ^2\right)]-\beta
   \left(\gamma +\omega ^2\right)^2 [y^2 \left(\gamma +\omega
   ^2\right)-5 \upsilon ^2] [17 y^2 \left(\gamma +\omega
   ^2\right)-5 \upsilon ^2]\nonumber\\&-&\left(\gamma +\omega ^2\right)^3 [y^2
   \left(\gamma +\omega ^2\right)-\upsilon ^2]^3]+\left(\gamma
   +\omega ^2\right)^2 [9 \mu  \zeta _{24} [\mu  \zeta
   _{24} \left(\gamma +\omega ^2\right)\nonumber\\&+&2 \upsilon  [\beta
   +\left(\gamma +\omega ^2\right) [\upsilon ^2+3 y^2 \left(\gamma
   +\omega ^2\right)]]]+6 \nu  y \zeta _{21}
   [\left(\gamma +\omega ^2\right) [3 \upsilon ^2\nonumber\\&+&y^2 \left(\gamma
   +\omega ^2\right)]-5 \beta ]-\nu ^2 \zeta
   _{21}^2]]{}^2].
\end{eqnarray}}Dynamics features of
3-rogue wave solutions are shown in Figs. 2-5. When $(\mu,\nu)=(0,
0)$, this is two high peaks and almost all energy of the rogue wave
is located on the two peaks in Fig. 2. when $(\mu,\nu)=(0, 100),
(100, 0), (100, 100)$, respectively, it is clearly that three rogue
waves break apart and form a set of three 1-rogue waves in Figs.
3-5.

\includegraphics[scale=0.55,bb=-20 270 10 10]{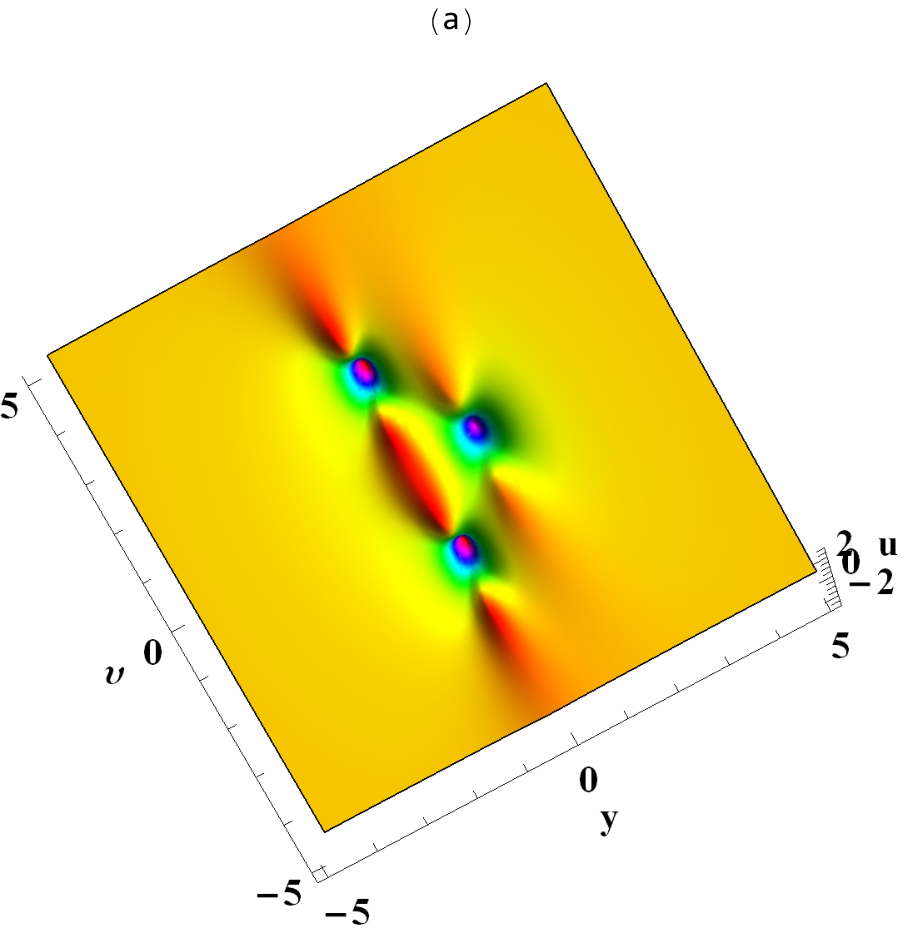}
\includegraphics[scale=0.45,bb=-270 310 10 10]{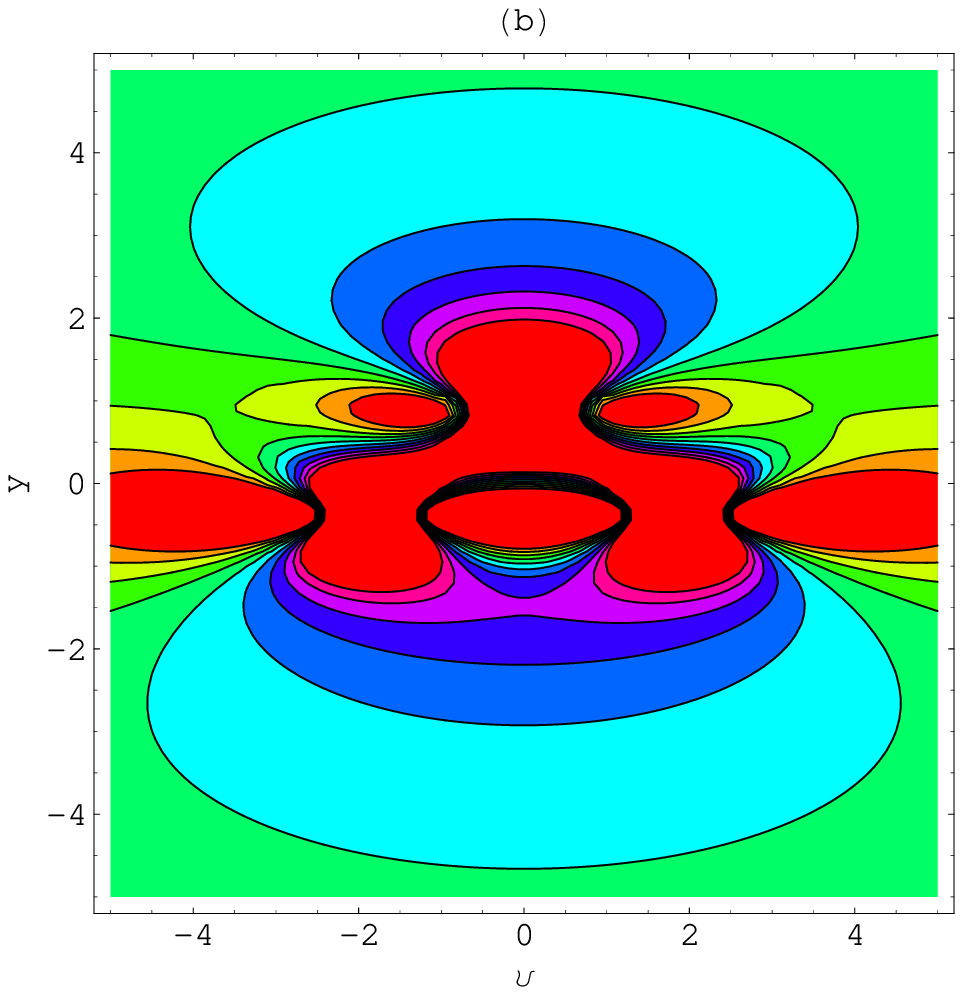}
\vspace{5.5cm}
\begin{tabbing}
\textbf{Fig. 3}. Rogue wave (13) with  $\alpha=12$, $\gamma=-8$,
$\beta=\omega=\zeta_{21}=\zeta_{24}=1$,\\ $\mu=0$, $\nu=100$, (a) 3D
graphic, (b) contour plot.
\end{tabbing}

\includegraphics[scale=0.55,bb=-20 270 10 10]{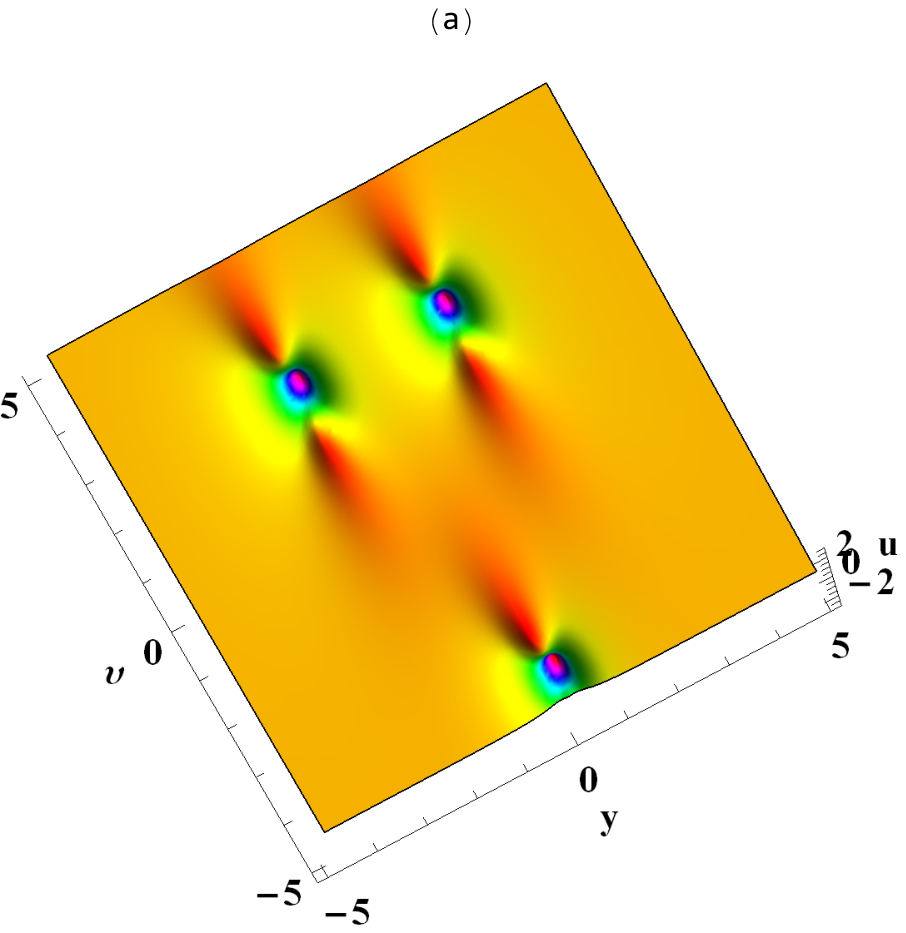}
\includegraphics[scale=0.45,bb=-270 310 10 10]{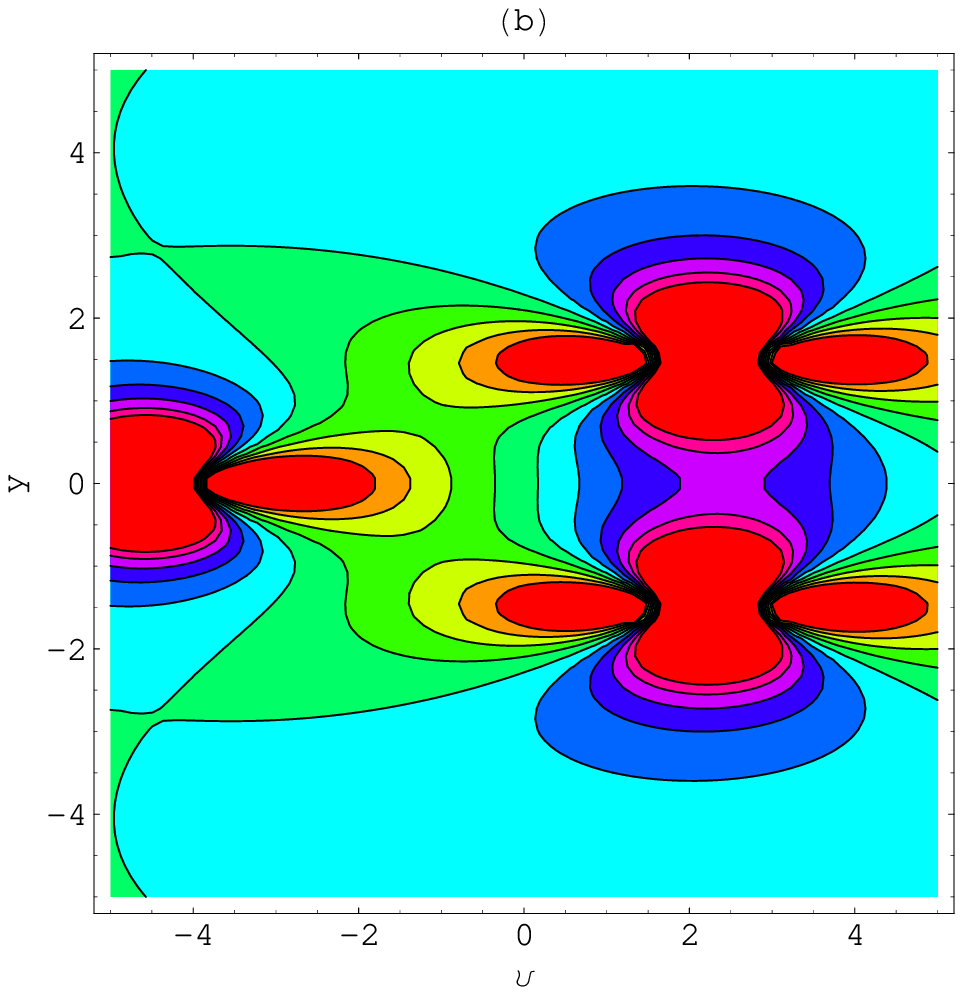}
\vspace{5.5cm}
\begin{tabbing}
\textbf{Fig. 4}. Rogue wave (13) with  $\alpha=12$, $\gamma=-8$,
$\beta=\omega=\zeta_{21}=\zeta_{24}=1$,\\ $\mu=100, \nu=0$, (a) 3D
graphic, (b) contour plot.
\end{tabbing}

\includegraphics[scale=0.55,bb=-20 270 10 10]{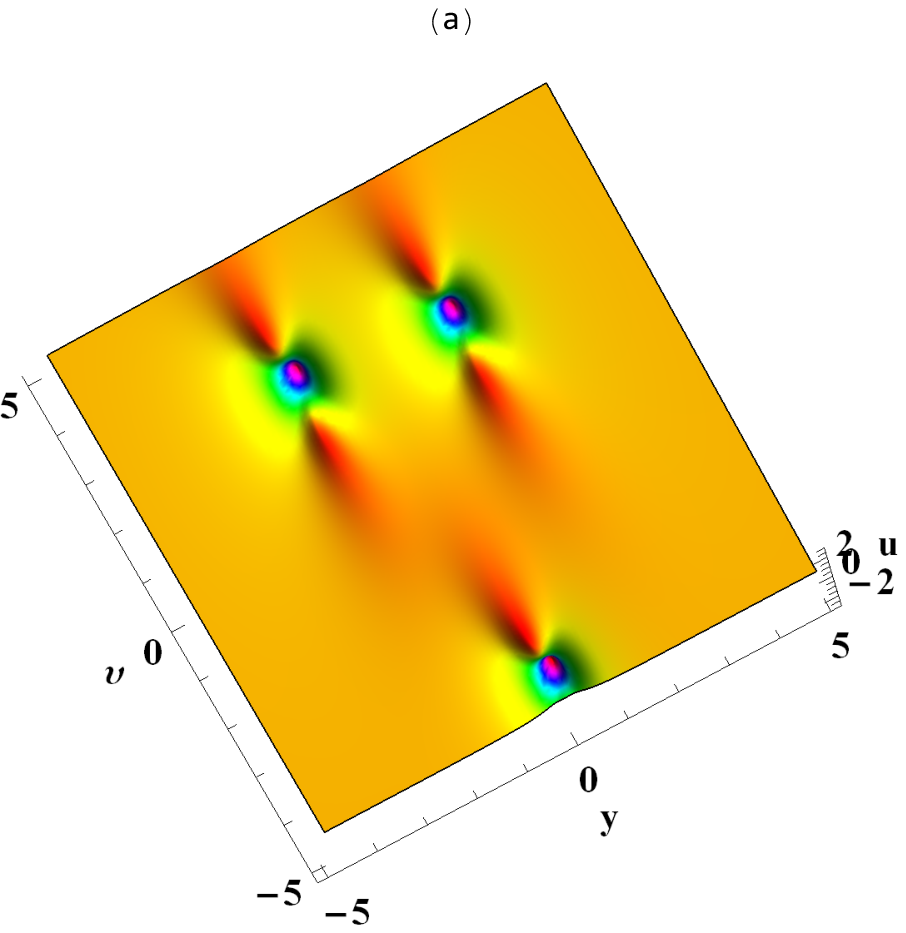}
\includegraphics[scale=0.45,bb=-270 310 10 10]{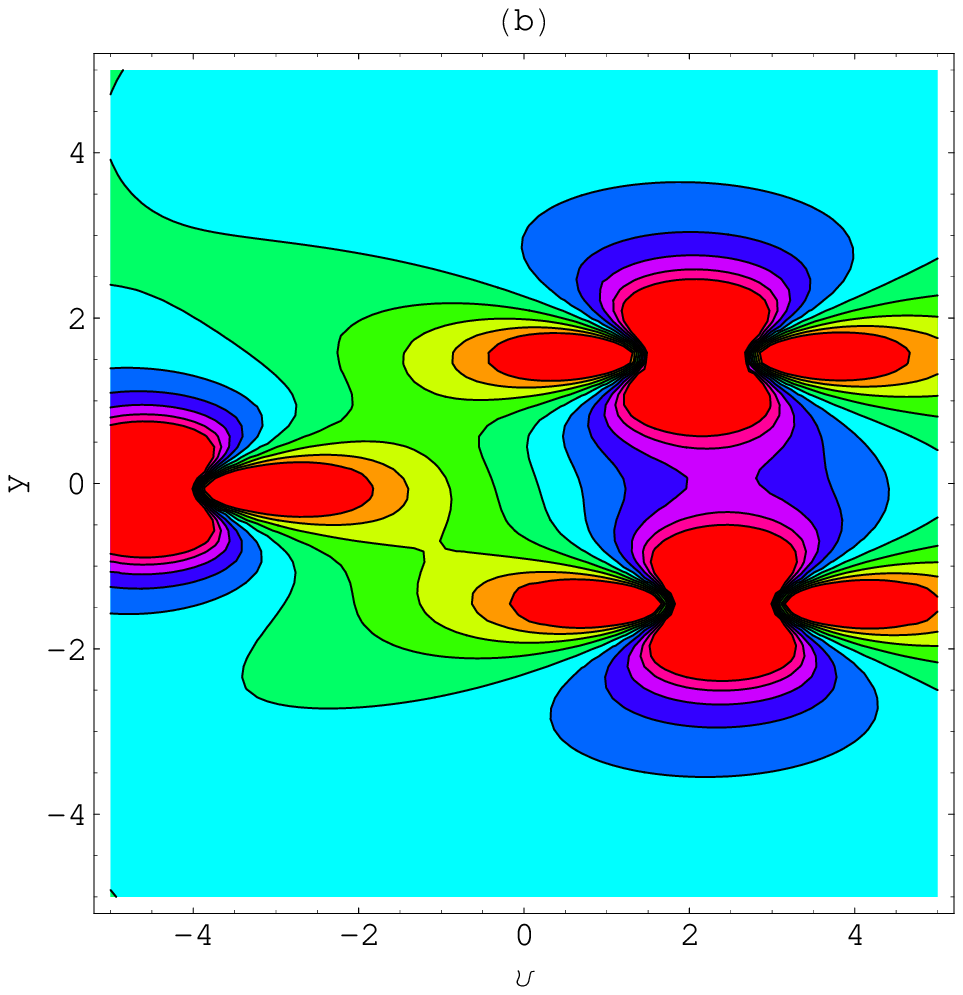}
\vspace{5.5cm}
\begin{tabbing}
\textbf{Fig. 5}. Rogue wave (13) with  $\alpha=12$, $\gamma=-8$,
$\beta=\omega=\zeta_{21}=\zeta_{24}=1$,\\ $\mu=\nu=100$, (a) 3D
graphic, (b) contour plot.
\end{tabbing}

\section{6-rogue wave solutions} \label{sec:2}
To search the 6-rogue wave solutions, we have {\begin{eqnarray}
\xi(\upsilon,y)&=&\upsilon ^{12}+y^8 \zeta _{48}+y^6 \zeta
   _{47}+y^4 \zeta _{46}+\upsilon ^{10}
   \left(y^2 \zeta _{26}+\zeta _{25}\right)\nonumber\\&+&y^2 \zeta _{45}+\upsilon ^8 \left(y^4
   \zeta _{29}+y^2 \zeta _{28}+\zeta _{27}\right)+2 \mu  \upsilon  [\upsilon ^6+y^6
   \zeta _{64}+y^4 \zeta _{63}\nonumber\\&+&\upsilon ^4 \left(y^2 \zeta _{69}+\zeta
   _{68}\right)+y^2 \zeta _{62}+\upsilon ^2 \left(y^4 \zeta _{67}+y^2 \zeta
   _{66}+\zeta _{65}\right)+\zeta _{61}]\nonumber\\&+&2 \nu  y [y^6+y^4 \left(\upsilon ^2
   \zeta _{57}+\zeta _{56}\right)+y^2 \left(\upsilon ^4 \zeta _{55}+\upsilon ^2
   \zeta _{54}+\zeta _{53}\right)+\upsilon ^6 \zeta _{60}\nonumber\\&+&\upsilon ^4 \zeta
   _{59}+\upsilon ^2 \zeta _{58}+\zeta _{52}]+\upsilon ^6 \left(y^6 \zeta
   _{33}+y^4 \zeta _{32}+y^2 \zeta _{31}+\zeta _{30}\right)\nonumber\\&+&\upsilon ^4 \left(y^8
   \zeta _{38}+y^6 \zeta _{37}+y^4 \zeta _{36}+y^2 \zeta _{35}+\zeta
   _{34}\right)+\upsilon ^2 (y^{10} \zeta _{44}+y^8 \zeta _{43}\nonumber\\&+&y^6 \zeta _{42}+y^4
   \zeta _{41}+y^2 \zeta _{40}+\zeta _{39})+\zeta _{51}+y^{12} \zeta _{50}+y^{10}
\zeta _{49}\nonumber\\&+&\left(\mu ^2+\nu ^2\right) [-\frac{3 \beta
}{\gamma +\omega ^2}+\upsilon ^2+y^2 \left(-\gamma -\omega
^2\right)],
\end{eqnarray}
}where $\zeta _i (i=25,\cdots, 69)$  is undetermined real constant.
Substituting Eq. (14) into Eq. (7) and equating the coefficients of
all powers $\upsilon$ and $y$ to zero, we obtain
\begin{eqnarray}
\zeta_{26}&=&-6 \left(\gamma +\omega ^2\right), \zeta_{29}=15
\left(\gamma +\omega ^2\right)^2, \zeta_{28}=690 \beta,\nonumber\\
\zeta_{33}&=&-20 \left(\gamma +\omega ^2\right)^3, \zeta_{32}=-1540
\beta \left(\gamma +\omega ^2\right),
\zeta_{31}=-\frac{18620 \beta ^2}{\gamma +\omega ^2},\nonumber\\
\zeta_{37}&=&1460 \beta  \left(\gamma +\omega ^2\right)^2,
\zeta_{36}=37450 \beta ^2, \zeta_{55}=-\frac{5}{\left(\gamma +\omega ^2\right)^2},\nonumber\\
\zeta_{38}&=&15 \left(\gamma +\omega ^2\right)^4,
\zeta_{35}=\frac{220500 \beta ^3}{\left(\gamma +\omega
^2\right)^2}, \zeta_{43}=-570 \beta  \left(\gamma +\omega ^2\right)^3,\nonumber\\
\zeta_{42}&=&-35420 \beta ^2 \left(\gamma +\omega ^2\right),
\zeta_{41}=\frac{14700 \beta ^3}{\gamma +\omega ^2}, \zeta_{52}=\frac{18865 \beta ^3}{3 \left(\gamma +\omega ^2\right)^6},\nonumber\\
\zeta_{54}&=&\frac{190 \beta }{\left(\gamma +\omega ^2\right)^3},
\zeta_{44}=-6 \left(\gamma +\omega ^2\right)^5,
\zeta_{57}=\frac{9}{\gamma +\omega ^2}, \zeta_{40}=-\frac{565950 \beta ^4}{\left(\gamma +\omega ^2\right)^3},\nonumber\\
\zeta_{50}&=&\left(\gamma +\omega ^2\right)^6, \zeta_{49}=58 \beta
\left(\gamma +\omega ^2\right)^4,
\zeta_{48}=4335 \beta ^2 \left(\gamma +\omega ^2\right)^2,\nonumber\\
\zeta_{64}&=&-5 \left(\gamma +\omega ^2\right)^3,
\zeta_{47}=\frac{798980 \beta ^3}{3},
\zeta_{27}=\frac{735 \beta ^2}{\left(\gamma +\omega ^2\right)^2},\nonumber\\
\zeta_{67}&=&-5 \left(\gamma +\omega ^2\right)^2,
\zeta_{63}=-45 \beta  \left(\gamma +\omega ^2\right), \zeta_{66}=-230 \beta,\nonumber\\
\zeta_{25}&=&-\frac{98 \beta }{\gamma +\omega ^2},
\zeta_{56}=-\frac{7 \beta }{\left(\gamma +\omega ^2\right)^2},
\zeta_{46}=\frac{16391725 \beta ^4}{3 \left(\gamma +\omega
^2\right)^2},\nonumber\\ \zeta_{53}&=&-\frac{245 \beta
^2}{\left(\gamma +\omega ^2\right)^4},
\zeta_{60}=-\frac{5}{\left(\gamma +\omega ^2\right)^3},
\zeta_{69}=9 \left(\gamma +\omega ^2\right),\nonumber\\
\zeta_{30}&=&-\frac{75460 \beta ^3}{3 \left(\gamma +\omega
^2\right)^3}, \zeta_{39}=\nu ^2 [-\frac{1}{\left(\gamma +\omega
^2\right)^7}-1]-\frac{159786550 \beta ^5}{3
\left(\gamma +\omega ^2\right)^5},\nonumber\\
\zeta_{34}&=&-\frac{5187875 \beta ^4}{3 \left(\gamma +\omega
^2\right)^4}, \zeta_{68}=-\frac{13 \beta }{\gamma +\omega ^2},
\zeta_{58}=\frac{665 \beta ^2}{\left(\gamma +\omega
^2\right)^5},\nonumber\\
\zeta_{45}&=&\frac{300896750 \beta ^5 \left(\gamma +\omega
^2\right)^2+3 \nu ^2 [\left(\gamma +\omega ^2\right)^7+1]}{3
\left(\gamma +\omega ^2\right)^6},
\zeta_{65}=-\frac{245 \beta ^2}{\left(\gamma +\omega ^2\right)^2},\nonumber\\
\zeta_{62}&=&-\frac{535 \beta ^2}{\gamma +\omega ^2},
\zeta_{61}=-\frac{12005 \beta ^3}{3 \left(\gamma +\omega ^2\right)^3},  \zeta_{59}=\frac{105 \beta }{\left(\gamma +\omega ^2\right)^4},\nonumber\\
\zeta_{51}&=&\frac{\beta  [878826025 \beta ^5 \left(\gamma +\omega
^2\right)^2+27 \nu ^2 [\left(\gamma +\omega ^2\right)^7+1]]}{9
\left(\gamma +\omega
   ^2\right)^8}.
\end{eqnarray}Substituting Eq. (14) and Eq. (15) into Eq. (6), the 6-rogue wave
solutions for Eq. (1) can be read as {\begin{eqnarray} u=\frac{6
\beta}{\alpha}
\left(\frac{\xi_{\upsilon\upsilon}}{\xi}-\frac{\xi_\upsilon^2}{\xi^2}\right),
\end{eqnarray}}where $\xi$ satisfies Eq. (14) and Eq. (15). Dynamics features of
3-rogue wave solutions are shown in Figs. 6-9. When $(\mu,\nu)=(0,
0)$, this is three high peaks and almost all energy of the rogue
wave is located on the two peaks in Fig. 9. when $(\mu,\nu)=(0,
1000), (1000, 0), (1000, 1000)$, respectively, it is clearly that
six rogue waves break apart and form a set of six 1-rogue waves in
Figs. 7-9.

\includegraphics[scale=0.55,bb=-20 270 10 10]{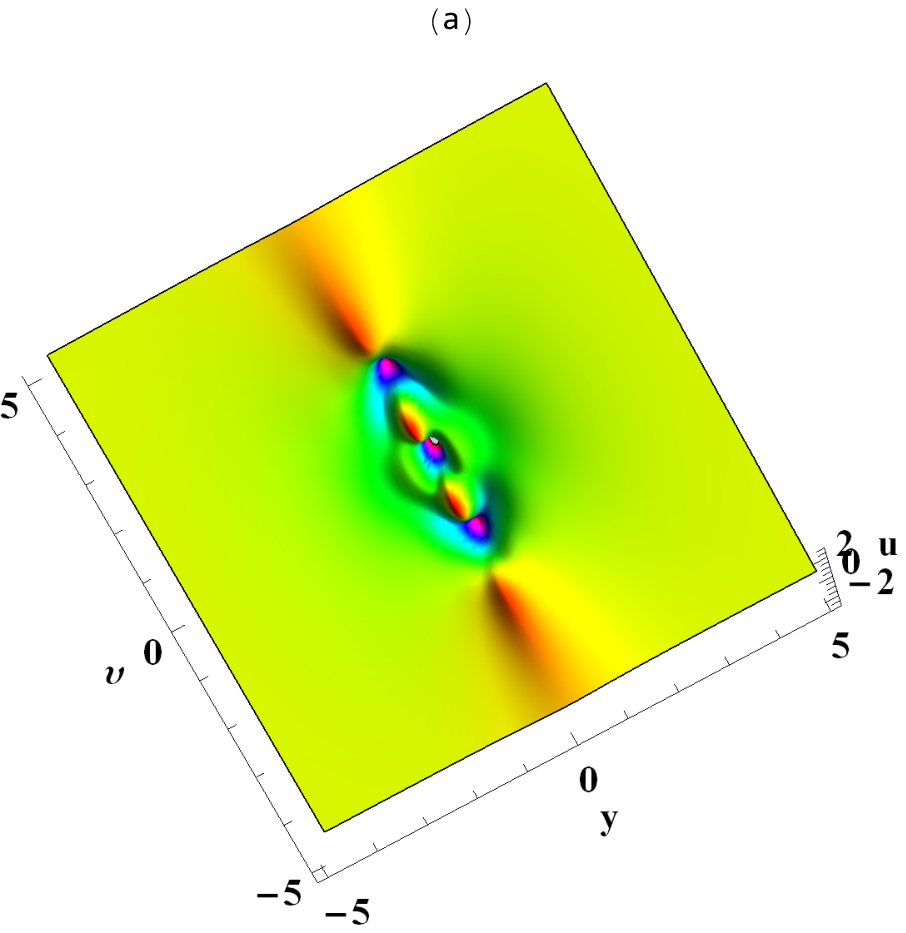}
\includegraphics[scale=0.45,bb=-270 310 10 10]{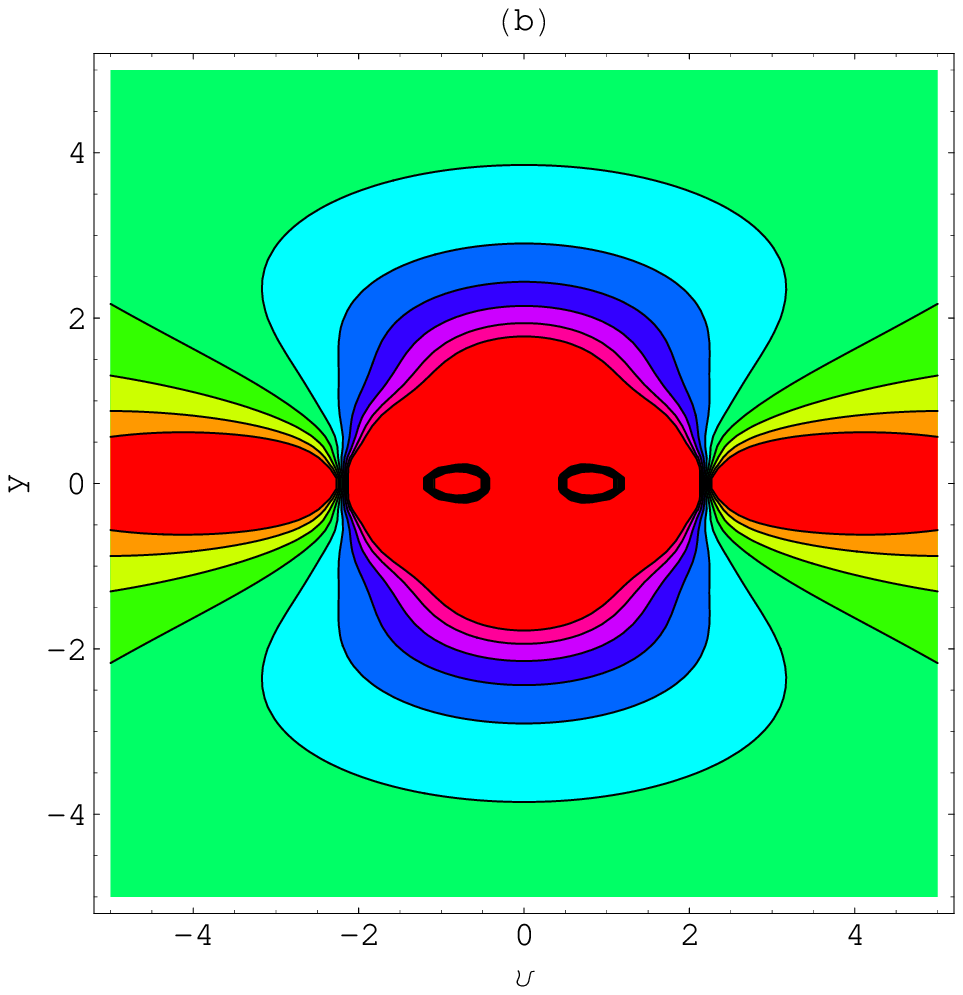}
\vspace{5.5cm}
\begin{tabbing}
\textbf{Fig. 6}. Rogue wave (16) with $\mu=\nu=0$, $\alpha=12$,
$\gamma=-8$, $\beta=\omega=1$,\\ (a) 3D graphic,  (b) contour plot.
\end{tabbing}

\includegraphics[scale=0.55,bb=-20 270 10 10]{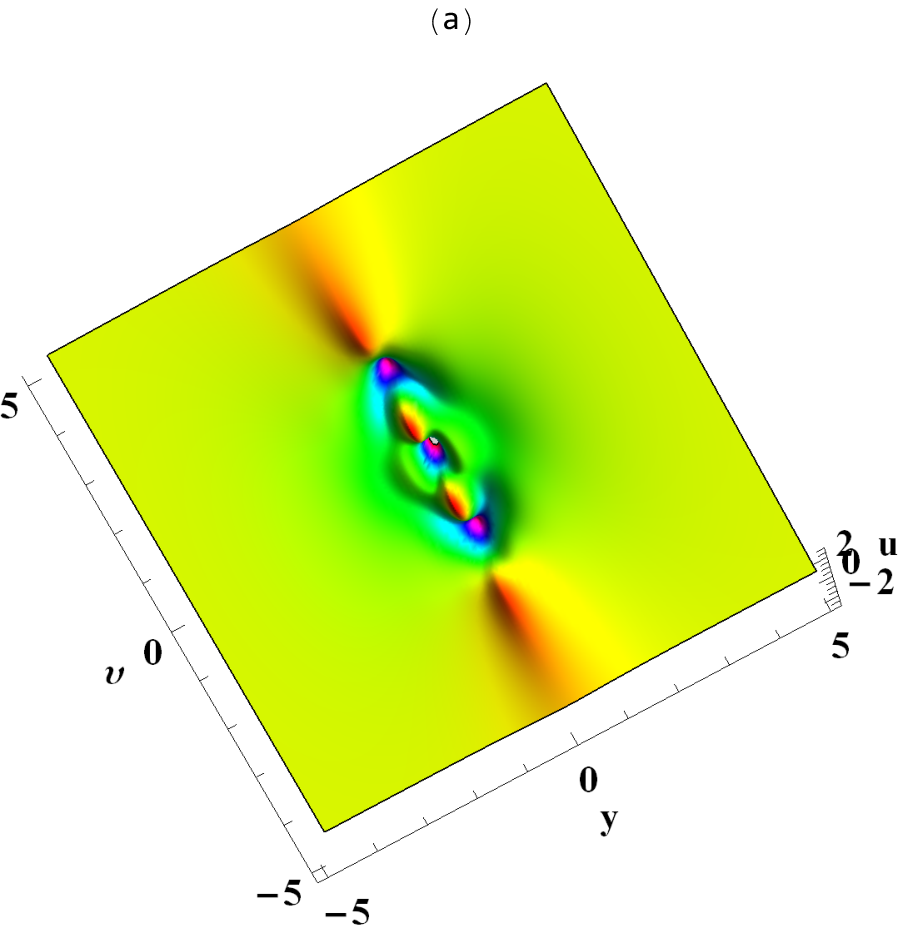}
\includegraphics[scale=0.45,bb=-270 310 10 10]{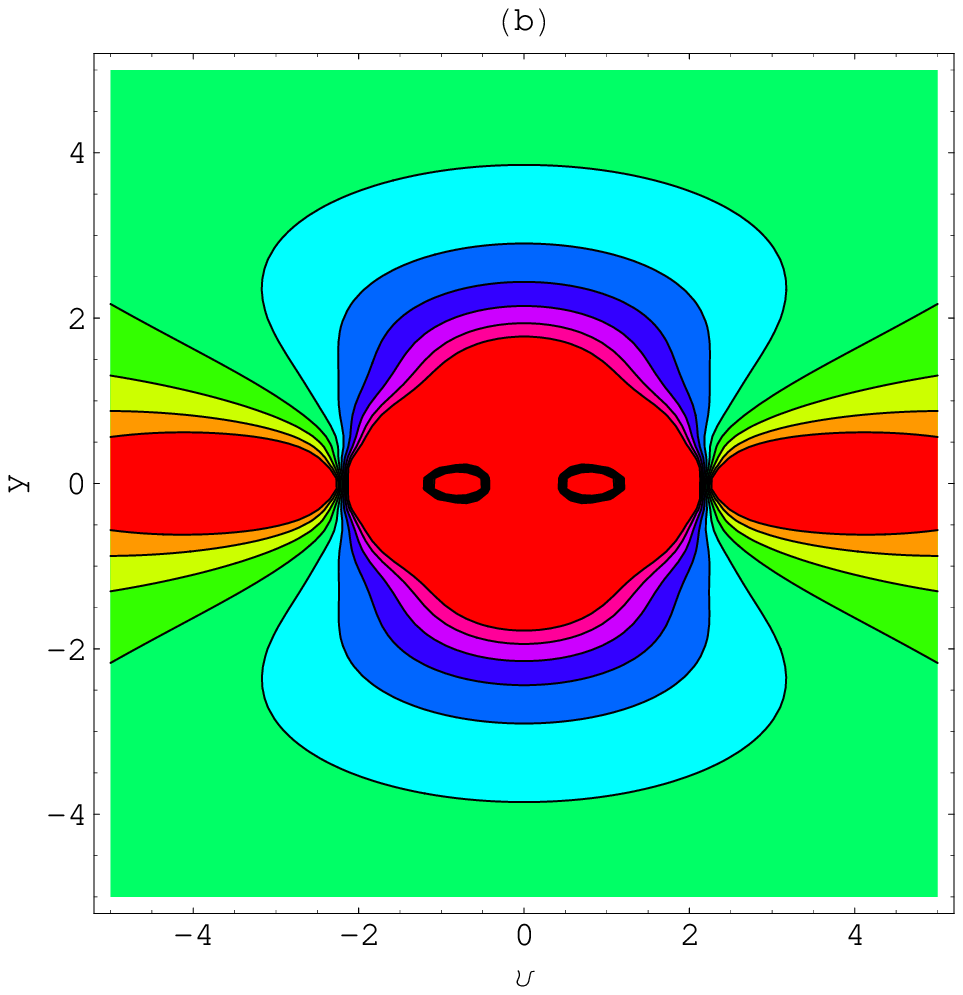}
\vspace{5.5cm}
\begin{tabbing}
\textbf{Fig. 7}. Rogue wave (16) with $\mu=0, \nu=1000$,
$\alpha=12$, $\gamma=-8$, $\beta=\omega=1$,\\ (a) 3D graphic,  (b)
contour plot.
\end{tabbing}

\includegraphics[scale=0.55,bb=-20 270 10 10]{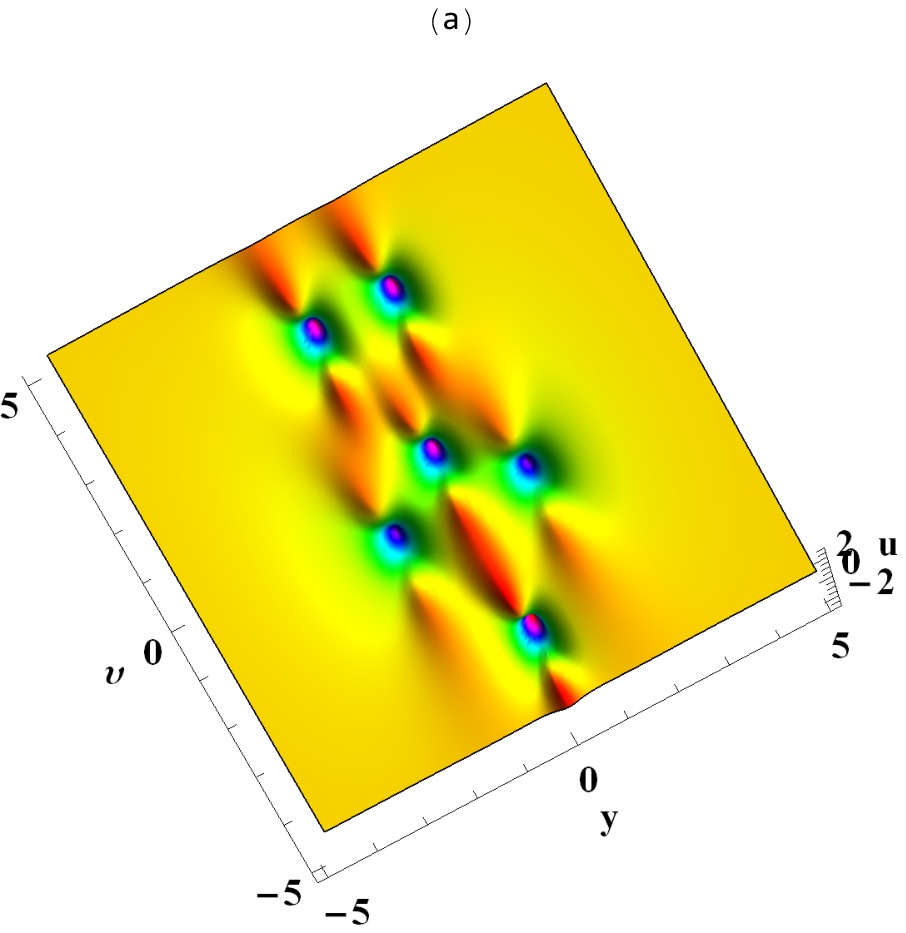}
\includegraphics[scale=0.45,bb=-270 310 10 10]{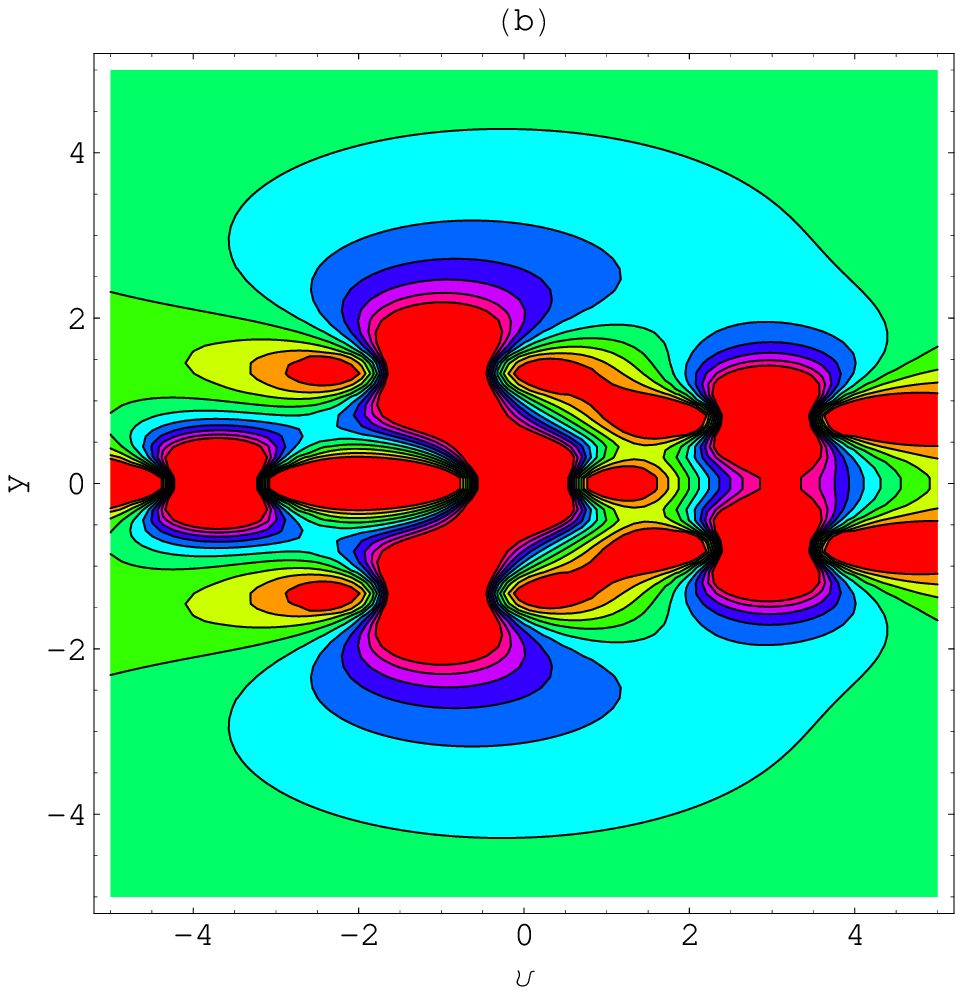}
\vspace{5.5cm}
\begin{tabbing}
\textbf{Fig. 8}. Rogue wave (16) with $\mu=1000, \nu=0$,
$\alpha=12$, $\gamma=-8$, $\beta=\omega=1$,\\ (a) 3D graphic,  (b)
contour plot.
\end{tabbing}

\includegraphics[scale=0.55,bb=-20 270 10 10]{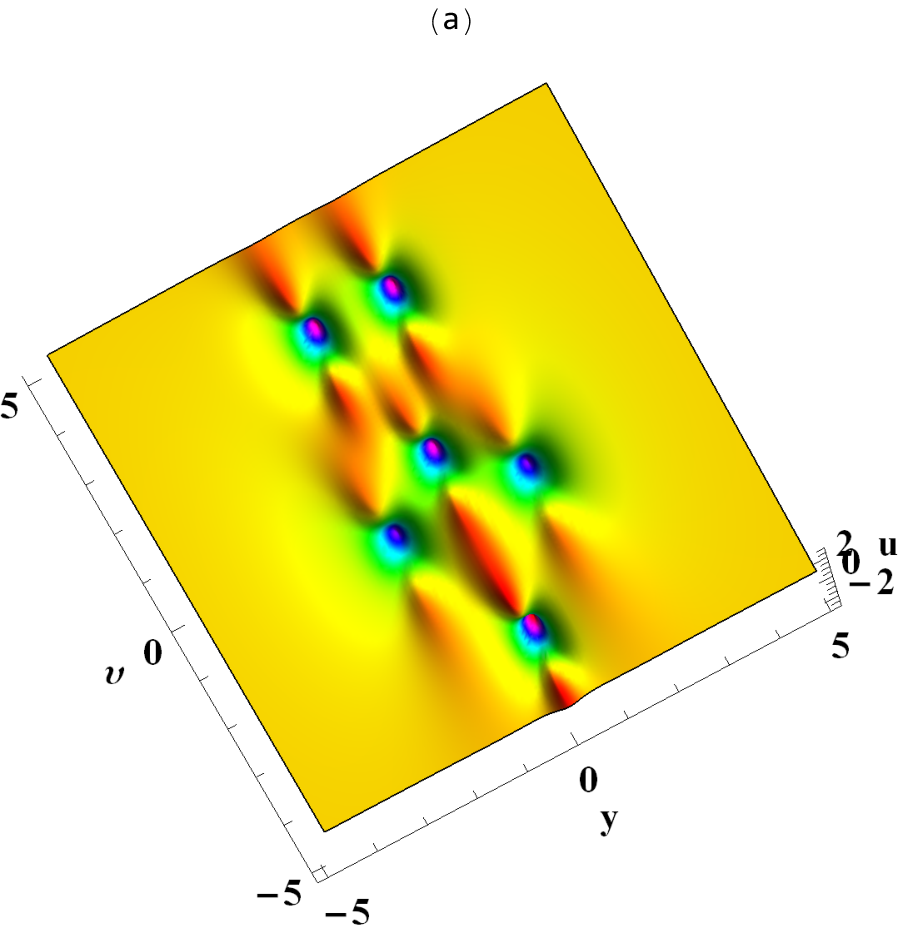}
\includegraphics[scale=0.45,bb=-270 310 10 10]{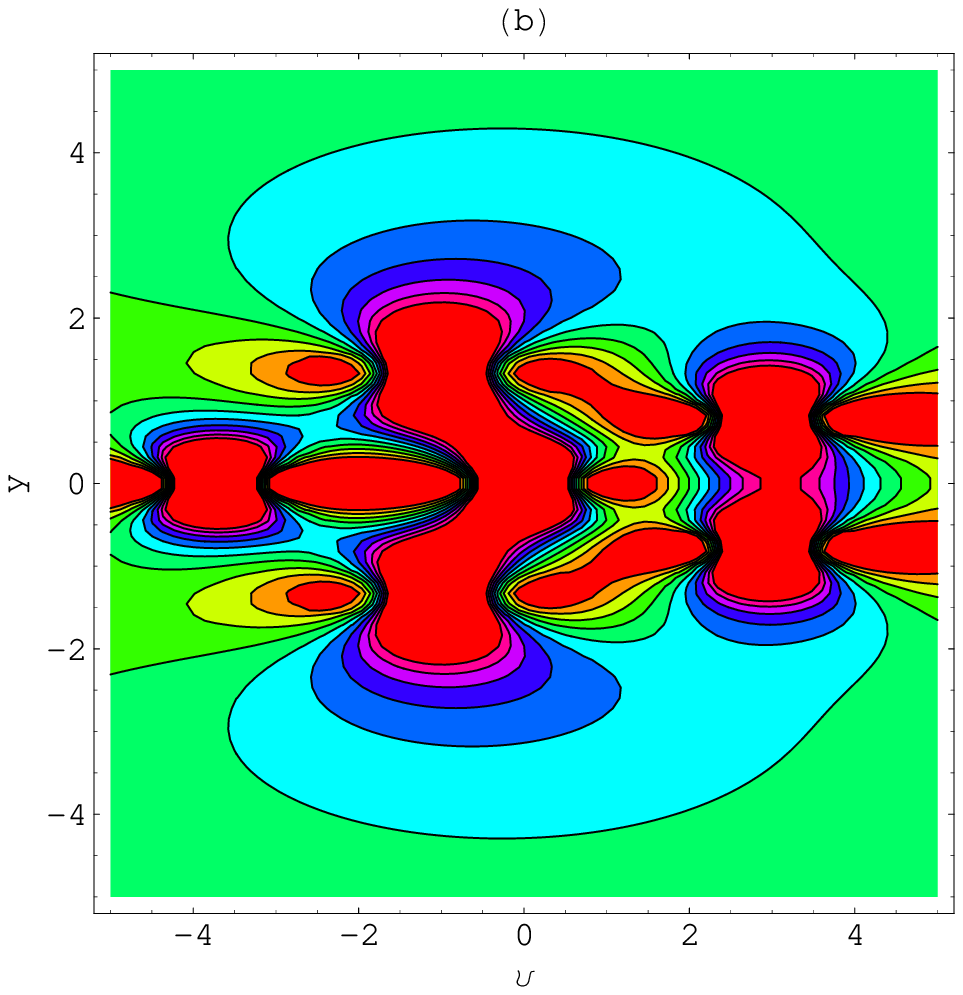}
\vspace{5.5cm}
\begin{tabbing}
\textbf{Fig. 9}. Rogue wave (16) with $\mu=\nu=1000$, $\alpha=12$,
$\gamma=-8$, $\beta=\omega=1$,\\ (a) 3D graphic,  (b) contour plot.
\end{tabbing}

\section{ Conclusion}
\label{sec:3} \quad In the paper,   a modified symbolic computation
approach is proposed. Compared with the original method, our method
does not need to find Hirota bilinear form. As an result, the
modified symbolic computation approach is applied to a generalized
(2+1)-dimensional Boussinesq equation. The 1-rogue wave solutions,
3-rogue wave solutions and 6-rogue wave solutions are obtained,
respectively. By selecting different values of $(\mu, \nu)$,
 their dynamics features are shown in Figs. 1-9. All the solutions have been put
 into the original equation by Mathematical software and verified to be correct.\\

\noindent {\bf Compliance with ethical standards}\\

\quad {\bf Conflict of interests} The authors declare that there is
no conflict of interests regarding the publication of this article.

\quad {\bf Ethical standard} The authors state that this research
complies with ethical standards. This research does not involve
either human participants or animals.


\begin{thebibliography}{}
%
%
\bibitem{Ref1}
Wazwaz, A.M.: Negative-order integrable modified KdV equations of
higher orders. Nonlinear Dyn., 93(3),  1371-1376 (2018).
\bibitem{s1}   Yin, Y.H., Ma, W.X., Liu, J.G., L\"{u}, X., Diversity of exact solutions to a (3+1)-dimensional
 nonlinear evolution equation and its reduction, Comput. Math. Appl., 76, 1275-1283 (2018).
\bibitem{s1}  Xu, G.Q.,  Wazwaz, A.M.: Characteristics of integrability, bidirectional solitons and
localized solutions for a (3 + 1)-dimensional generalized breaking
soliton equation. Nonlinear Dyn.,  96, 1989-2000 (2019).
\bibitem{s2} Wazwaz, A.M., Kaur, L.: New integrable Boussinesq equations of distinct
dimensions with diverse variety of soliton solutions. Nonlinear
Dyn., 97(1), 83-94 (2019).
\bibitem{s3} Lan, Z.Z.: Rogue wave solutions for a coupled nonlinear Schr?dinger equation in the birefringent optical
fiber. Appl. Math. Lett., 98, 128-134 (2019).
\bibitem{s4} Gao, L.N., Zhao, X.Y., Zi, Y.Y., Yu, J., L\"{u}, X.: Resonant behavior of multiple wave solutions to
a Hirota bilinear equation. Comput. Math. Appl., 72, 1225-1229
(2016).
\bibitem{s1} Wazwaz, A.M.: The integrable time-dependent sine-Gordon equation with multiple optical kink solutions.
Optik, 182, 605-610  (2019).
\bibitem{s5} Xu, T., Chen, Y.: Semirational solutions to the coupled Fokas¨CLenells
equations. Nonlinear Dyn.,  95, 87-99 (2019)
\bibitem{s6} Osman, M.S.,  Wazwaz, A.M.: An efficient algorithm to construct multi-soliton
rational solutions of the (2+ 1)-dimensional KdV equation with
variable coefficients. Appl. Math. Comput., 321,  282-289 (2018).
\bibitem{s7} Li, Y.Z., Liu, J.G.: New periodic solitary wave solutions for the
new (2+1)-dimensional Korteweg-de Vries equation. Nonlinear Dyn.,
91(1), 497-504 (2018).
\bibitem{s11} Ahmed, I., Seadawy, A.R., Lu, D.C.: Kinky breathers, W-shaped and multi-peak solitons interaction
in (2 + 1)-dimensional nonlinear Schr\"{o}dinger equation with Kerr
law of nonlinearity. Eur. Phys. J. Plus,  134, 120 (2019).
\bibitem{s12} Ghanbari, B., Inc, M.,  Rada, L., Solitary wave
solutions to the tzitzeica type equations obtained by a new
efficient approach, J. Appl. Anal. Comput., 9(2),  568-589 (2019)
\bibitem{s13} X.Y. Xie, G.Q. Meng, Dark solitons for the (2+1)-dimensional Davey-Stewartson-like
equations in the electrostatic wave packets, Nonlinear Dyn. 93
(2018) 779-783.
\bibitem{s14} Clarkson, P.A.,  Dowie, E.: Rational solutions of the Boussinesq
equation and applications to rogue waves. Trans. Math. Appl., 1(1),
 tnx003 (2017).
\bibitem{s15} Su, J.J., Gao, Y.T., Ding, C.C.: Darboux transformations and rogue wave solutions of a generalized AB system for the
geophysical flows. Appl. Math. Lett., 88, 201-208 (2019).
\bibitem{s16} Wang, X.B., Han, B., Tian, S.F.: General coupled nonlinear Schr\"{o}dinger equation: Breather
waves and rogue waves on a soliton background, and dynamics.
Superlattice. Microst., 128, 83-91 (2019).
\bibitem{s19} Liu, J.G., You, M.X., Zhou, L., Ai, G.P.: solitary wave, rogue wave and periodic solutions for the
(3 + 1)-dimensional soliton equation.  Z. Angew. Math. Phys.,
 70(1), 4 (2019).
\bibitem{s20} Yang, B., Chen, Y.: Dynamics of rogue waves in the partially PT-symmetric nonlocal Davey-Stewartson
systems. Commun. Nonlinear Sci. Numer. Simulat., 69,  287-303
(2019).
\bibitem{s21} Zha,Q.L.: A symbolic computation approach to constructing rogue waves with a controllable center in the nonlinear
systems. Comput. Math. Appl., 75(9),  3331-3342 (2018).
\bibitem{s18} Liu, Y.K., Li, B., An, H.L.: General high-order breathers, lumps in the
(2+1)-dimensional Boussinesq equation. Nonlinear Dyn., 92, 2061-2076
(2018).
\bibitem{s23} Chen, M.: Exact solutions of various Boussinesq systems.
Appl. Math. Lett., 11, 45-49 (1998).
\bibitem{s18} Korpel, A., Banerjee, P.: Proc. IEEE 72, 1109-1130
(1984).
\bibitem{s18} Fan, E.: Integrable System and Computer Algebra. Science
Press, Beijing (2004).
\bibitem{s18} Figen, \"{O}., Haci, B.,  Hasan, B.: On the complex and hyperbolic structures for
 the (2 + 1)-dimensional boussinesq water equation. Entropy, 17(12), 8267-8277 (2015).
\bibitem{s18} Wang, X.B., Tian, S.F., Qin, C.Y., et al.: Characteristics of
the breathers, rogue waves and solitary waves in a generalized (2 +
1)-dimensional Boussinesq equation. Euro. Phys. Lett., 115(1), 10002
(2016).
\bibitem{s18} Ma, H.C., Deng, A.P.: Lump solution of (2+1)-dimensional
Boussinesq equation. Commun. Theor. Phys., 65, 546-552 (2016).
\bibitem{s18} Liu, W.H., Zhang, Y.F.:  Multiple rogue wave solutions of the (3+1)-dimensional
Kadomtsev-Petviashvili-Boussinesq equation. Z. Angew. Math. Phys.,
 70, 112 (2019).
\bibitem{s18} Liu, W.H., Zhang, Y.F.: Multiple rogue wave solutions for a (3+1)-dimensional Hirota
bilinear equation. Appl. Math. Lett., 98,  184-190 (2019).
\bibitem{s26}  Gaillard, P.: Rational solutions to the KPI equation and multi rogue waves. Ann. Phys., 367,  1-5 (2016).
\bibitem{s29} Zhao, Z.L., He, L.C., Gao, Y.B.: Rogue Wave and Multiple Lump Solutions of
the (2+1)-Dimensional Benjamin-Ono Equation in Fluid Mechanics.
Complexity,, 2019, 8249635 (2019).
\bibitem{s29} Zhao, Z.L., He, L.C.: Multiple lump solutions of the (3+1)-dimensional potential
Yu-Toda-Sasa-Fukuyama equation. Appl. Math. Lett., 95, 114-121
(2019).


\end{thebibliography}


\end{document}